# Photochemical generation of E' centre from Si-H in amorphous $SiO_2$ under pulsed ultraviolet laser radiation.

F Messina and M Cannas

*Dipartimento di Scienze Fisiche ed Astronomiche dell'Università di Palermo*

*Via Archirafi 36, I-90123 Palermo, Italy*

email: fmessina@fisica.unipa.it

**Abstract**

*In situ* optical absorption spectroscopy was used to study the generation of E' centres ($\equiv$Si•) in amorphous $SiO_2$ occurring by photo-induced breaking of Si-H groups under 4.7eV pulsed laser radiation. The dependence from laser intensity of the defect generation rate is consistent with a two-photon mechanism for Si-H rupture, while the growth and the saturation of the defects are conditioned by their concurrent annealing due to reaction with mobile hydrogen arising from the same precursor. A rate equation is proposed to model the kinetics of the defects and tested on experimental data.

The E' centre, consisting in un unpaired electron on a threefold coordinated silicon atom ($O_3\equiv Si\bullet$), is one of the fundamental point defects in amorphous silica (a-$SiO_2$); the generation of this centre upon exposure of silica to laser radiation is often the main cause of transmission loss of the material in the ultraviolet (UV) range, due to its absorption band peaked at 5.8eV [1-2]. Many papers have been devoted to study the generation of E' under laser radiation, pointing out a variety of possible mechanisms, such as conversion of oxygen-deficient centres, strained bonds or Si-bonded impurities, or non-radiative decay of self-trapped excitons [2,3-11].

One of the most common precursors of the E' centre is the Si-H group, from which the defect can be generated by a photolysis process under UV laser photons [5-7,10-11]:

$$\text{Si-H} + h\nu \rightarrow \text{E'} + H_0 \qquad (1)$$

E' centres produced by (1) are transient, at least at room temperature. In fact, hydrogen atoms $H_0$ coming as well by (1), dimerize in $H_2$ which diffuses and recombines with E' [8,10-15]:

$$\text{E'} + H_2 \rightarrow \text{Si-H} + H_0 \qquad (2)$$

If E' centres have been generated from Si-H by a pulsed laser irradiation, after the laser is switched off reaction (2) causes the decay in a typical time scale of a few hours of almost all the defects initially produced [14-15]. Moreover, the reaction is expected to occur also during the irradiation session, mainly in the time interval between a pulse and the successive one (interpulse); hence, even the growth kinetics of the defects is conditioned by the interplay between pulse-induced generation (1) and interpulse annealing (2). Due to its characteristics, only a limited amount of data is available on process (1), whose comprehensive study requires *in situ* measurements, widely available only since a few years [9-11,15-16]. *In situ* measurements are suitable to monitor the transient kinetics of E' and also the most appropriate to find out if the UV-induced Si-H breaking is a single- or a two-photon process.

The main purpose of this work is to study generation of E' centres by UV-induced breaking of Si-H in a system where this process can be isolated; a material of choice to this aim is wet fused quartz, where in a recent experiment founded on *in situ* optical absorption (OA) measurements it was demonstrated that formation of E' centres under 4.7eV pulsed laser radiation is due to reaction (1) [15]. Moreover, it was found that this material contains no dissolved $H_2$ prior to laser exposure and there are no other OA signals significantly overlapping with the 5.8eV band. Hence, the generation of E' from Si-H can be isolated and studied selectively.

We point out that the information derived here may be relevant also for the understanding of other systems, such as Si/SiO$_2$ interfaces, where Si-H breaking is a very common process which results in the formation of P$_b$-type interface centers (whose structure is Si$_3$≡Si• at the (111)Si/SiO$_2$ interface), this being an important degradation mechanism of microelectronics devices [17-18]. In particular, under laser radiation, Si-H breaking at Si/SiO$_2$ interfaces has been observed to occur either by a photothermal mechanism, or by direct photolysis [19].

The experiments reported here were performed on Herasil 1 $a$-SiO$_2$ samples, 5×5×1 mm$^3$ sized, OH content ~200 ppm and optically polished on all surfaces. The presence of Si-H groups was checked by preliminary Raman measurements showing their typical 2250cm$^{-1}$ signal [20-21]. Si-H concentration was roughly estimated to be ~5×10$^{17}$cm$^{-3}$.

Samples were irradiated at room temperature perpendicularly to their minor surface by the IV harmonic of the pulsed radiation emitted by a Q-switched Nd:YAG laser ($\tau$=5 ns duration, $\Delta t$=1s interpulse, pulse energy h$\nu$≈4.7 eV). The laser beam was checked to have a uniform intensity profile over a (6.0±0.1)mm diameter circular spot; then, pulse energy was measured by a pyroelectric detector. Finally, the ratio of pulse energy to the beam section and duration gives laser peak intensity $\Gamma$. We verified that during laser irradiation the temperature of the samples does not vary significantly from T$_0$=300K.

We performed 8 irradiation sessions on different virgin samples at different laser peak intensities, from (2.6±0.2)×10$^6$ W/cm$^2$ to (19±1)×10$^6$ W/cm$^2$, each session consisting in a few thousand pulses. We measured *in situ* the absorption profile of the sample in the UV, once during each interpulse, by a single-beam AVANTES S2000 optical fibre spectrophotometer system, based on a CCD detector and working in the 200-400nm spectral range. The difference OA profiles measured after different numbers of pulses N during an irradiation session (figure 1) shows that laser radiation induces the growth of the 5.8eV band of the E'-centre [1-2]; the defect is absent in the as-grown material as checked by the absence of its paramagnetic signal in preliminary ESR measurements. The 5.8eV band grows during irradiation without significant width variations. The small negative component at ~5.1eV is due to the bleaching of the Ge-related B$_{2\beta}$ band [14-15].

From the peak 5.8eV absorption coefficient and the known absorption cross section [22], we calculated the concentration [E'] of the defects, which is plotted in figure 2 as a function of irradiation fluence (F=Γ×τ×N) for 3 representative irradiation sessions. We see that the kinetics show a saturation tendency above ~50Jcm$^{-2}$, but the saturated concentration [E']$_S$ grows with laser intensity. We note that also the initial slope d[E']/dF(F=0) of the curves grows with Γ. The initial slope may be determined by a linear fit in the first ~3Jcm$^{-2}$; hence, the so-obtained values are converted to generation rates per pulse, given by R= d[E']/dN=Γ×τ×d[E']/dF; R is plotted in figure 3 versus Γ. By a least-square fit of these data with the function R=aΓ$^b$ (the continuous line in the plot) we obtain b=2.2±0.2; this means that the behaviour of R is consistent with a quadratic dependence from Γ.

As discussed in the introduction, for this material exposed to 4.7eV laser light E' centres are transient since they arise from breaking of Si-H precursors [14-15]. Indeed, as soon as the laser is switched off, we observe a progressive decay of the centres, bringing their concentration almost to zero in a few hours. As an example, in the inset of figure 1 we show the kinetics of [E'] comprising also the first ~10$^3$ s of the post-irradiation decay stage, for a sample irradiated with 1200 laser pulses of Γ=12×10$^6$Wcm$^{-2}$ peak intensity (the kinetics is plotted as a function of time). The post-irradiation decay was dealt with in previous papers [14-15]; from now on, we will focus only on the generation stage of the process, which is the main interest of this paper.

Data in figure 3 permit to address an important feature of the laser-induced breaking process of Si-H (reaction (1)). Indeed, the quadratic dependence of R on peak intensity demonstrates that two-photon processes are involved in E' generation. We note that the simple observation (figure 2) that the kinetics start with different slopes allows to rule out a simple one-photon mechanism, which would lead to a constant d[E']/dF(F=0). A two-photon mechanism is consistent with experimental and simulative data on Si-H group: this centre does not show any measurable OA at energies below silica bandgap, and its lowest transition is predicted at ~9eV, leading to an anti-bonding state [6, 23]. Hence, the simplest E' generation mechanism consistent with present results is two photon absorption (TPA) by Si-H leading to the excited state with consequent breaking of the bond. Yet, other nonlinear processes are conceivable, such as production of excitons by TPA followed by non-radiative decay on Si-H. Finally, we should consider the possibility of extrinsic impurities in the material playing some role in E' generation, since we evidenced that in synthetic

SiO$_2$ irradiated in the same conditions E' are not generated; impurities may for example serve as intermediate states in two step absorption assisting hole-electron pairs generation [24].

Several works in literature have discussed the generation mechanisms of E' in silica under laser radiation, distinguishing between single- and multi-photon processes [3-7,9-11,25-26], but most of them dealt with permanent defects, and only a few have directly observed *in situ* the transient E' centres originated from the Si-H precursor [10-11,25]. Only in [10] and [25], the dependence from laser energy density of the initial generation rate of transient E' centres was studied in synthetic SiO$_2$ exposed to KrF (5.0eV) or ArF (6.4eV) laser radiation, and reported to be respectively linear and sublinear. These results are in disagreement with the quadratic dependence found here. At the moment the reason of this discrepancy is unclear, but it may be related to differences in the materials employed, leading to the activation of different mechanisms leading to Si-H breaking, or to the coexistence of (transient) E' centres not arising from Si-H.

Finally, it may be interesting to compare present results with what is known on laser-induced Si-H breaking on Si/SiO$_2$ interfaces. In these systems, theoretical work has fixed the bonding-nonbonding electronic transition of Si-H to be at 8.5 eV [27], not far from the 9eV value found in a-SiO$_2$; consistently, Si-H photolysis was observed to occur by single-photon absorption of F$_2$ laser (7.9eV) radiation [19]. On the other side, Si-H breaking under less energetic photons efficiently occurs by a photothermal mechanism peculiar of the Si/SiO$_2$ system, in which the Si-H bond is broken by providing the ~2.6eV dissociation energy by heating due to strong absorption of the laser light by the silicon substrate [19, 28-29].

We proceed now to address the issue of quantitatively modelling the growth kinetics of the defects. The kinetics of E' (figure 2) show a saturation at high fluences: this fact cannot be due to exhaustion of the Si-H precursor, whose native concentration estimated from Raman measurements is more than one order of magnitude higher than [E']$_S$. Moreover, [E']$_S$ is found to depend on $\Gamma$, whereas saturation by precursor exhaustion would fix [E']$_S$ to a value independent from irradiation conditions. Then, the saturation must be due to an equilibrium between generation and a concurrent depletion mechanism of the induced E' population. Due to the behaviour of E' in the post-irradiation stage, we argue that reaction (2) is effective also in the interpulse time span, and is a possible candidate for the depletion mechanism.

In this scheme, the kinetics of [E'] on the scale of many laser pulses (figure 2) should be described by the following rate equation, provided that N is approximately treated as a continuous variable:

$$\frac{d[E']}{dN} \approx \frac{\Delta[E']}{\Delta N} \approx R - 2k_0[E'][H_2]\Delta t = R - k_0\Delta t[E']^2 \qquad (3)$$

where R is the generation rate, which equals the initial growth slope of the kinetics (figure 3), while the negative term accounts for the decrease of [E'] during the $\Delta t$=1s interpulse due to reaction (2), calculated within the stationary-state approximation[a] for $H_0$ (factor 2 derives from each $H_2$ passivating two E') [14,30]. For the last equality we have used the relation: $[H_2]$=1/2[E'], which derives from $H_0$ and E' being generated together and from the absence of dissolved $H_2$ prior to laser exposure [15]. The reaction constant $k_0$ between E' and $H_2$ at room temperature had already been measured to be $k_0$=(8.3±0.8)×$10^{-20}$cm$^3$s$^{-1}$ [15].

We found that Eq. (3) results to be in disagreement with the experimental kinetics. In particular, the saturation concentration of E', found from (3) when d[E']/dN=0, is given by $[E']_{SP}$=$(R/k_o\Delta t)^{1/2}$; for instance we obtain $[E']_{SP}$ =2.0×$10^{16}$cm$^{-3}$ for the kinetics at $\Gamma$=12×$10^6$Wcm$^{-2}$, which results higher than the actual $[E']_S$ =1.2×$10^{16}$cm$^{-3}$ (figure 2); a similar situation is found for all the kinetics. Moreover, we verified that the agreement with experimental data is not improved by taking into account the statistical distribution of the diffusion parameters of $H_2$ related to the amorphous structure of the silica matrix [14,16]. This implies that annealing driven by $H_2$-diffusion (reaction (2)) significantly slows down the growth of the defects but, if alone, is insufficient to explain the observed saturation concentrations, which are still lower than predicted; for this reason, an additional negative term has to be added to the rate equation.

On this basis, we found empirically that all the kinetics can be well fitted by adding a linear term to (3) as follows:

$$\frac{d[E']}{dN} = R(1-\alpha[E']) - k_0\Delta t[E']^2 \qquad (4)$$

this being equivalent to supposing a (linear) concentration-dependence of the generation rate. In detail, best-fit curves of figure 2 were obtained by fitting to experimental data solutions of eq. (4), depending on the free parameter $\alpha$. The values of $\alpha$ all fall in the interval (5.7±1.4)×$10^{-17}$cm$^3$.

We propose now a simple qualitative interpretation of the linear term, based on a more accurate discussion of the generation process (1): in general, a $H_0$ produced by (1) diffuses in the matrix and may experience two different fates: it can either meet another $H_0$ and dimerize in $H_2$ or come across an E' and passivate it; for this reason, apart from slow annealing due to reaction (2), the E' centres undergo a much faster decay (FD) due to recombination with a portion of the $H_0$ population made available by each pulse. Moreover, the portion of $H_0$ involved in the FD is expected to increase with E' concentration, which enhances the probability of encountering an E' before meeting another $H_0$. A FD stage with similar features was directly observed *in situ* for non bridging oxygen hole centres (NBOHC) produced by $F_2$ laser irradiation at T=300K, and occurs on a typical time scale which is shorter than the 1s interpulse [16]. Since the FD cannot be directly observed with the time resolution available here, it is incorporated *de facto* in the generation term, which actually must be interpreted as the net concentration of E' generated by (1) and surviving fast recombination with $H_0$[b]. Now, given that the portion of $H_0$ quickly recombinating with E' must increase with E' concentration as the irradiation session progresses, we expect a consequent reduction of the generation rate from its initial value R, which in first approximation can be represented by a linear term in [E'], i.e. $-\alpha$[E'] in (4). We acknowledge that measures resolved on a time scale shorter than the interpulse are required to confirm this interpretation by a thorough investigation of the dynamics involving fast atomic hydrogen diffusion; however, we point out that the main advantages of the simple model proposed here are that it permits to understand the kinetics and the saturation of E' as a consequence of hydrogen-related reactions, and it is able to reproduce independent datasets coming from several irradiation sessions with only one free parameter, whose origin can be understood on a qualitative basis.

---

[a] The stationary-state approximation for $H_0$ automatically takes into account both reaction (2) and the reaction between E' and $H_0$ made available at the right side of (2)

In conclusion, we studied the generation of E' centres from Si-H groups in a-$SiO_2$ under pulsed 4.7eV laser radiation. The dependence of the initial generation rate from laser intensity is quadratic, demonstrating a two-photon mechanism for E' generation. The kinetics and the saturation of the process are the result of the competition between the action of radiation and the annealing of E' due to reaction with hydrogen arising from the same precursor. On this basis, a rate equation model was proposed and tested against experimental data.

**Acknowledgements**

---

[b] It is worth to note that the FD due to partial recombination with the transient $H_0$ population available just after each pulse is not taken into account in the -2k[E'][$H_2$] term, which is founded on the stationary state approximation for $H_0$, and so takes into account only $H_0$ produced by (2)

The authors thank Prof. Boscaino and group in Palermo for support and enlightening discussions, Y. Ouerdane and K. Médjahdi from Université J. Monnet in Saint Etienne (France) for Raman measurements and G. Napoli, G. Lapis for technical support.

**Figure captions**

**Figure 1:** OA measured *in situ* after different numbers of 4.7eV laser pulses with $\Gamma=12\times10^6 Wcm^{-2}$ peak intensity. Inset: growth and decay of E' concentration, as calculated from the peak of the 5.8eV OA.

**Figure 2**: Kinetics of [E'] (open symbols) during 3 irradiation sessions performed with different laser intensities. The continuous lines are least-square fits with analytical solutions of equation (5), with $\alpha$ as free fitting parameter.

**Figure 3:** E' initial generation rate per pulse against laser intensity. The line is a least-square fit of the data with the function $y=ax^b$, yielding $b=2.2\pm0.2$.

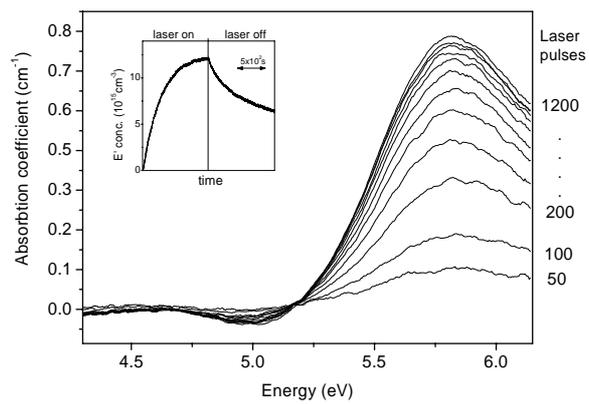

**FIGURE 1**

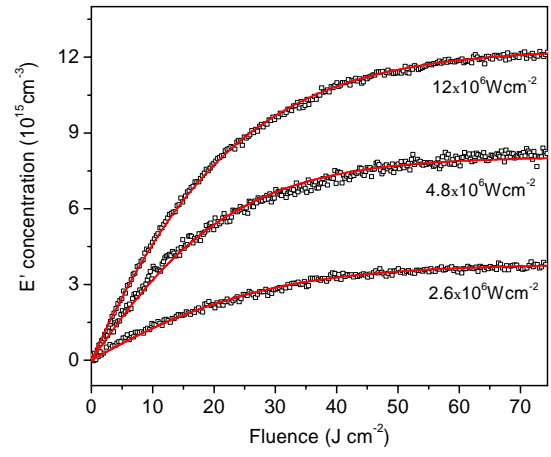

**FIGURE 2**

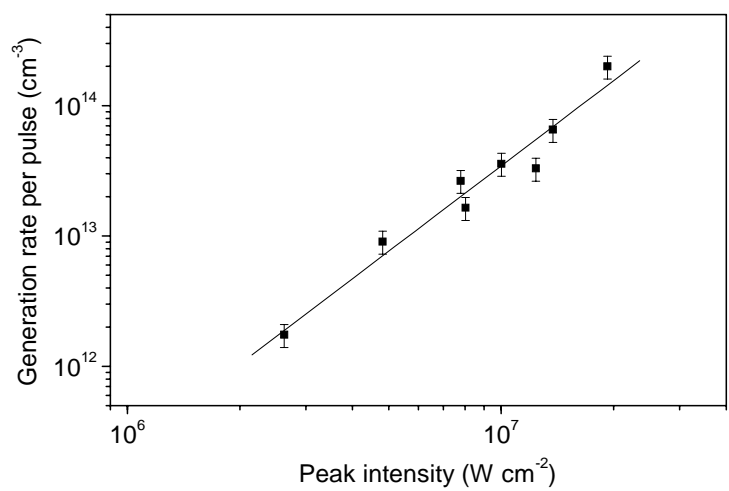

**FIGURE 3**